\documentclass[12pt,a4paper]{article}

\usepackage{times} 
\usepackage[a4paper, lmargin=2cm, rmargin=2cm , tmargin=2.5cm , bmargin=2.25cm  ,footskip=1.25cm]{geometry} 
\usepackage{graphicx} 
\usepackage[square,sort,comma,numbers]{natbib} 
\usepackage{amsmath} 
\usepackage[hyphens]{url} 
\usepackage{soul}
\usepackage{fancyhdr} 
\usepackage[explicit]{titlesec} 
\usepackage{abstract} 
\usepackage{authblk} 
\usepackage{wrapfig}
\usepackage{tabularx}
\usepackage{multirow} 
\usepackage{multicol}
\usepackage{colortbl} 
\usepackage{booktabs}
\usepackage{array}
\usepackage[super]{nth} 
\usepackage{algorithm}
\usepackage[end]{algpseudocode}
\usepackage{enumitem}
\usepackage{subfig}
\usepackage{array}
\usepackage[pdftex]{hyperref}
\usepackage{comment}
\usepackage{bm}

\newcommand\Tstrut{\rule{0pt}{2.6ex}}         
\newcommand\Bstrut{\rule[-0.9ex]{0pt}{0pt}}   

\newcommand{\matlab}{\textsc{Matlab }}
\newcommand{\simulink}{\textsc{Simulink }}
\titleformat{\section}{\normalfont\bfseries}{\thesection}{1em}{\MakeUppercase{#1}}  

\titleformat{\subsection}{\normalfont\bfseries\small}{\thesubsection}{1em}{{#1}} 

\titleformat{\subsubsection}{\normalfont\small}{\thesubsubsection}{1em}{{#1}} 

\algrenewcommand\algorithmicend{\textbf{end}}
\algrenewtext{EndFor}{\algorithmicend}

\makeatletter
\newcommand{\thickhline}{ \noalign {\ifnum 0=`} \fi \hrule height 1pt \futurelet \reserved@a \@xhline }
\newcommand{\morethickhline}{ \noalign {\ifnum 0=`} \fi \hrule height 2pt \futurelet \reserved@a \@xhline }
\newcolumntype{"}{@{\hskip\tabcolsep\vrule width 1pt\hskip\tabcolsep}}
\makeatother

\newlength{\Oldarrayrulewidth}

\begin{document}

\title{\normalsize\normalfont\bfseries \MakeUppercase {Worst-Case Pointing Performance Analysis for Large Flexible Spacecraft}}
\date{}
\author[(1)]{COURIE Ilham}
\author[(1)]{SANFEDINO Francesco}
\author[(1)]{ALAZARD Daniel}
\affil[(1)]{ \textit{ISAE-SUPAERO, Université de Toulouse, 31055 Toulouse Cedex 4, France (e-mail: ilham.courie@student.isae-supaero.fr, francesco.sanfedino@isae.fr, daniel.alazard@isae.fr)}}

\renewcommand\Authands{, }
\renewcommand{\Authfont}{\normalsize\normalfont \bfseries}
\renewcommand{\Affilfont}{\normalsize\normalfont}
\renewcommand{\abstractnamefont}{\bfseries\normalsize\MakeUppercase} 

\maketitle

\begin{abstract}
This paper presents a tool, \textsc{Pelib}, developed in \matlab/\simulink environment to perform pointing performance analysis based on European pointing standards. \textsc{Pelib} is designed as an extension of the Satellite Dynamics Toolbox (SDT), which derives the Linear Fractional Transformation (LFT) models of flexible space structures. The addition of \textsc{Pelib} will allow the users of SDT to perform pointing performance analysis of real mission scenarios in the same environment used for control synthesis. \textsc{Pelib} offers as well the possibility to take into account uncertainties in the system. This feature represents an enhancement to the current verification tools available in the European space industry community by providing the worst-case pointing budget.
The capabilities of \textsc{Pelib} were demonstrated in a case study involving a spacecraft model with two flexible solar arrays. Several error sources as well as uncertain parameters were included in this model. The nominal performance has been investigated using \textsc{Pelib} and compared with the current European reference tool. The worst-case performance is also investigated with the new feature of \textsc{Pelib} to obtain the worst-case performance budget.
\end{abstract}

\thispagestyle{fancy}

\section{Introduction}

Pointing performance budgeting is a very crucial part of the design phase of a spacecraft for the current and future space missions, especially for scientific and Earth observation applications that tend to have very demanding control performances which can be easily affected by various disturbances. The disturbances in a spacecraft mission are caused by various internal and external sources such as solar pressure, gravity gradient, reaction wheels of the Attitude and Orbit Control System (AOCS), Solar Array Driving Mechanism (SADM) as well as sensors noise magnified by the attitude control law. Indeed, These disturbances and their impacts to the pointing performance need to be investigated as this information in early design phase can be very important for deriving the control performance requirements as well as to prevent additional design iterations due to pointing performance in more advanced phases.

With the publication of ECSS Control performance guidelines \cite{european_cooperation_for_space_standardization_ecss-e-hb-60-10a_2010} and ECSS Control Performance \cite{european_cooperation_for_space_standardization_ecss-e-st-60-10c_2008} by the European Cooperation for Space Standardization (ECSS), the pointing budget verification process can now be carried out with standardized rules and a clear mathematical basis. Furthermore, ESA also released in 2011 the Pointing Error Engineering Handbook (PEEH) \cite{essb-hb-e-003_working_group_esa_2011} that provides the methodology and the engineering framework that cover the step-by-step process of pointing performance budgeting with guidelines, examples and recommendations that are consistent with ECSS documents.

In the past, some efforts have been done to develop computer-based tool to perform error budget automatically. For instance, Astos Solutions, under ESA contract, has developed the Pointing Error Engineering Tool (PEET) \cite{m_casasco_g_saavedra_s_weikert_j_eggert_m_hirth_t_ott_h_su_pointing_2013}, a \matlab-based tool for pointing error analysis in-line with ESA standard \cite{european_cooperation_for_space_standardization_ecss-e-st-60-10c_2008,essb-hb-e-003_working_group_esa_2011}. ESA has used this software for pointing error analysis in several space missions \cite{m_casasco_g_saavedra_s_weikert_j_eggert_m_hirth_t_ott_h_su_pointing_2013} such as Euclid, MetOp-SG, and Proba-3.

However, the use of this tool is limited to the analysis of nominal plants, since uncertainties cannot be taken into account in a generalized Linear Fractional Transformation (LFT) model. The verification and validation (V\&V) process then consists in a time-consuming non-global Monte-Carlo approach, that risks to miss rare but possible worst-case scenarios.

This paper provides a novel tool in \matlab and \simulink environment to perform pointing error budgeting called Pointing Error Library (\textsc{Pelib}) that aims to fill this gap. \textsc{Pelib} which consists of various Matlab functions is a natural extension of the Satellite Dynamics Toolbox (SDT) \cite{d_alazard_c_cumer_k_h_tantawi_linear_2008, daniel_alazard_satellite_nodate,daniel_alazard_francesco_sanfedino_satellite_2020, sanfedino2021integrated, sanfe2019, d_alazard_f_sanfedino_satellite_2021} already implemented by the authors in the past under ESA, CNES, ONERA, Thales, and Airbus DS contracts for modelling of complex flexible space systems in minimal LFT form. Each function in this library follows the standard and methodology described in ECSS and ESA Handbook from the classification of error sources to the evaluation of pointing error budget. In addition to this, a simple and compact graphical user interface (GUI) based on Matlab App Designer is also provided to define the pointing systems and requirements. Furthermore, the authors also propose a novel methodology based on robust control and analysis in order to extend the current approach in ESA Handbook to uncertain parametric systems which will be described in section \ref{sec:extension}. This feature allows the user to obtain worst-case pointing budget in terms of worst-case gain, worst-case variance, and worst-case low frequency (DC) gain.

Finally, a case study is proposed to further illustrate the capability of this tool in section \ref{sec:application}. In this case study, the pointing performance of a spacecraft model fitted with two flexible solar panels and two SADMs is investigated. Validation of the nominal case will be obtained by comparing the results with PEET. The analysis is then extended to include various uncertain parameters in the model.

\section{Methodology}
\label{sec:method}

The pointing error engineering process as described by ESA handbook covers all the steps from defining the requirements to obtaining the pointing error budget from all the error sources. This methodology involves four systematic analysis steps (AST) shortly summarized in the following section.

\subsection{Pointing Error Requirements} \label{subsec:requirements}

\begin{figure}[h!]
  \centering
  \includegraphics[width=0.7\columnwidth]{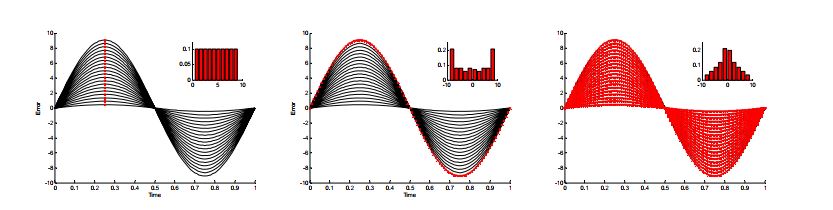}
  \caption[Statistical interpretation]{Statistical interpretation : ensemble(left), temporal(middle), mixed(right) \cite{european_cooperation_for_space_standardization_ecss-e-st-60-10c_2008}}
  \label{fig:SI}
\end{figure}

Before starting the analysis, it is necessary to formulate the requirements. Pointing error requirement is defined as the specification of probability that the system output of interest does not deviate more than a given amount from the target output with a level of confidence \begin{math}P_{c}\end{math} \cite{essb-hb-e-003_working_group_esa_2011}. As described in ESA handbook, the following parameters need to be specified to define unambiguous pointing error requirements: maximum error value \begin{math}e_{r}\end{math}, statistical interpretation, pointing error metric in term of performance/knowledge and window time, and the level of confidence.

The statistical interpretation as discussed in \cite{european_cooperation_for_space_standardization_ecss-e-st-60-10c_2008} describes how the pointing errors are defined in the domain of evaluation. There are three statistical interpretations generally used: ensemble, temporal and mixed (see Fig. \ref{fig:SI}). Ensemble interpretation is linked to the statistics of worst-case error that happen in all realizations regardless their time of occurrence (e.g maximum error for each observation during a satellite lifetime). Temporal interpretation is linked to the time domain behavior of error with worst-case realization. Lastly, mixed interpretation takes into accounts the statistics for both ensemble and temporal domain.

\subsection{Error Sources Characterization (AST-1)} \label{subsec:pesdef}
A pointing error can be defined as the response of the system to internal or external physical phenomena referred as pointing error sources (PES) that affect the performance of the system \cite{essb-hb-e-003_working_group_esa_2011}. In order to perform the pointing performance budgeting, it is necessary to know and properly model the behavior of each PES. Hence, The first step for this analysis (AST-1) is the identification of the error sources which involves classification and characterization of contributing errors. Following the guideline provided in ESA handbook \cite{essb-hb-e-003_working_group_esa_2011}, the pointing error sources in \textsc{Pelib} are described by random variable and random process \cite{js_bendat_and_ag_piersol_random_2010}. Currently, \textsc{Pelib} covers all type of PES described in ESA handbook including: time constant PES for error which is constant in time but can vary in its ensemble of realization (for example misalignment or biases); time-random random variable described by their probability density function (PDF) which can further be subjected to ensemble randomness in their statistical parameter; stationary Ergodic random process defined by power spectral density (PSD). Furthermore, \textsc{Pelib} also supports specific type of PES such as drift, periodic, gyro and star sensor noises. The periodic error source is modelled as random variable with bimodal distribution based on the sinusoidal amplitude and frequency. Moreover, the amplitude can also be an ensemble random parameter.

\subsection{Transfer Analysis (AST-2)}
Transfer analysis (AST-2) transforms the PES into pointing error contributor (PEC) and \textsc{Pelib} only considers transfer analysis through Linear Time-Invariant (LTI) systems and using frequency domain approach described in ESA handbook. The frequency domain approach consists in analyzing the transfer between some inputs and outputs of the system as extensively discussed in \cite{js_bendat_and_ag_piersol_random_2010}. When a PES is transferred through a stable and strictly proper LTI transfer function $\mathbf{H}(\mathrm{i}2\pi \omega)$, where $\omega$ is the frequency, the output PEC can be determined. According to the ESA handbook \cite{essb-hb-e-003_working_group_esa_2011}, If the input PES is a time-constant random variable than the output PEC can be determined by using the system low frequency (DC) gain. For a time-random random variable PES with unknown error spectrum then the system \begin{math}H_{\infty}\end{math}-norm will be used as an upper bound approximation. For time-random random process PES, the input PSD ($G_{ee}$) will be transformed into output PSD ($G_{yy}$) based on the relation
\begin{align}
\begin{split}
\begin{array}{c@{\qquad}c}
    G_{yy}(\omega) = |\mathbf{H}(\mathrm{i}2\pi \omega)|^2 G_{ee}, \mathrm{for \, SISO}
    \\
    {G}_{yy}(\omega) = \mathbf{H}(\mathrm{i}2\pi \omega) G_{ee} H^*(\mathrm{i}2\pi \omega), \mathrm{for \, MIMO}
    \label{eq:ast2}
\end{array}
\end{split}
\end{align}

with \begin{math} \mathbf{H}^*(\mathrm{i}2\pi \omega)=\mathbf{H}^T(-\mathrm{i}2\pi \omega) \end{math} is the complex conjugate transpose. The output variance will be evaluated using the following equation 
\begin{equation}
    \sigma_{PEC}^2 =  \int_{0}^{\infty} G_{yy}(\omega) d\omega
    \label{eq:sigmapec}
\end{equation} 
which corresponds to RMS-norms \begin{math}\left\|e_c\right\|_{rms}\end{math} described in \cite{s_boyd_c_barrat_linear_1991}.

\subsection{Error Index Contribution (AST-3)}

\begin{wrapfigure}[15]{r}{0.5\columnwidth}
\centering
\includegraphics[width=0.5\columnwidth]{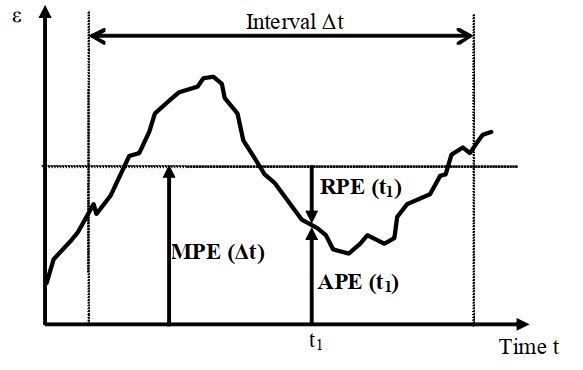}
\caption[Example of APE, MPE and RPE pointing error metrics]{Example of APE, MPE and RPE pointing error metrics \cite{european_cooperation_for_space_standardization_ecss-e-st-60-10c_2008}}
\label{fig:metric}
\end{wrapfigure}
Error index contribution (AST-3) evaluates the error within a time window by applying pointing error metrics as part of the requirements in section \ref{subsec:requirements}.

The definition of each error metric is discussed comprehensively in \cite{essb-hb-e-003_working_group_esa_2011}. For time-random PES of random variable type, the pointing error metrics will be applied based on the tables B-2 to B-7 in \cite{european_cooperation_for_space_standardization_ecss-e-hb-60-10a_2010}. For time-random PES of random process type, index specific PSD weighting functions \begin{math}F_{metric}\end{math} derived in \cite{m_e_pittelkau_pointing_2003,rl_lucke_sw_sirlin_am_san_martin_new_1992,d_s_bayard_state-space_2003} are more appropriate. In order to perform LTI analysis, rational approximation \begin{math}\Tilde{F}_{metric}\end{math} of the weighting functions are given in table 10-3 in \cite{essb-hb-e-003_working_group_esa_2011} such that \begin{math}F_{metric}(\omega) \Tilde{=} \Tilde{F}_{metric}(i2\pi \omega)\end{math}. Then, the variance of the output can be described by the following equation:
\begin{equation}
    \sigma_{metric}^2 =  \int_{0}^{\infty}\Tilde{F}_{metric}\,G_{ee}(\omega) d\omega
    \label{eq:sigmarp}
\end{equation}  

\subsection{Error Evaluation (AST-4)}
The final step of the buget analysis is to evaluate the contribution of the error of each type to obtain the overall budget with a given level of confidence. According to \cite{essb-hb-e-003_working_group_esa_2011}, there are two methods that can be used to evaluate the error contribution : simplified and advanced statistical method. 

\subsubsection{Simplified statistical method}

The simplified method evaluates the total error based on the mean and variance of each contributing error. This method assumes the applicability of central limit theorem \cite{js_bendat_and_ag_piersol_random_2010}. In simplified method, the errors will be summed based on the summation rules provided in ESA handbook: The means $\mu_{i}$ are summed linearly and the standard deviations $\sigma_{i}$ are summed either quadratically (uncorrelated) or linearly (fully correlated) and the total error $\epsilon_{index}$ is calculated using the following equation:
\begin{equation}
    \varepsilon_{index} = |\mu_{index}| + n_p . \sigma_{index}
    \label{eq:totalerror}
\end{equation}
with \begin{math}n_p\end{math} being the confidence factor that complies with given level of confidence \begin{math}P_c\end{math} requirement for Gaussian distributions. This method is exact or provides accurate results when the PES is dominated of Gaussian type error. For the case where the errors are mostly non-Gaussian, then the simplified method can lead to a large systematic error since central limit theorem is no longer applicable as discussed in \cite{m_hirth_h_su_t_ott_m_casasco_g_ortega_peet_2017}.

\subsubsection{Advanced statistical method}

The advanced statistical method or \textit{exact error combination} as referred in \cite{european_cooperation_for_space_standardization_ecss-e-hb-60-10a_2010} evaluates the total error based on the PDF of each error source and their joint PDF. Analytically, the PDF for the summation of independent random variables, for example $z = x + y$, with $x$ and $y$ independent random variables with PDF $p_{X}$ and $p_{Y}$ respectively, is given by the convolutional integral :
\begin{equation}
    p_{Z}(z) = \int p_{X}(x) p_{Y}(z-x) dx = \int p_{Y}(y) p_{X}(z-y) dy
    \label{eq:jointpdf}
\end{equation}
However, generally the closed form solution of this integral is difficult to obtain or does not exist. Even with numerical integration, the information of the joint PDF is generally not known especially for summation of random variables with different distributions. Alternatively. a sample-based approach can be used where random samples with given PDF for each PES are generated and summed linearly. The total error PDF which is the convolution of each error as well as the Cumulative Distribution Function (CDF) can be then easily obtained by directly using the generated data. Such method is also used in PEET and it was also further discussed in \cite{m_hirth_h_su_et_al_PEET_2016}
that the loss of accuracy compared to the analytical approach (see Eq. \eqref{eq:jointpdf}) can be reduced by choosing sufficiently large sample size, for example less than 1\% of error can be achieved with sample size of 1 million. Finally the total error budget which corresponds to the given level of confidence can be obtained by taking the inverse of the CDF. This can be easily achieved by interpolation of the CDF data at the demanded level of confidence. Both simplified and advanced statistical method are available in \textsc{Pelib}.

\section{Extension to uncertain systems}
\label{sec:extension}
\subsection{Worst-case variance and gain}
Considering a system $\mathbf H(\mathrm s,\boldsymbol\Theta)$, where $s$ is the Laplace variable, depending on the uncertain parameter vector $\boldsymbol\Theta$ varying in a given parametric domain $\mathcal D_{\boldsymbol \Theta}$, the objective is to compute the worst-case standard deviation $\sigma_{wc}$ (or the wost-case $H_2$ norm):
\[
\sigma_{wc}=\max_{\boldsymbol\Theta \in \mathcal D_{\boldsymbol \Theta}}\|\mathbf H(\mathrm s,\boldsymbol\Theta)\|_2\;.
\]
It is assumed that $\mathbf H$ is strictly proper and stable for all $\boldsymbol\Theta$ in $\mathcal D_{\boldsymbol \Theta}$.

Since there is no \matlab  function to compute $\sigma_{wc}$, an heuristic method, based on the function \texttt{systune}, is proposed. Indeed \texttt{systune} is quite efficient to solve robust performance control design problems \cite{APKA2015}. The worst-case analysis problem is then turned into a parametric robust control design problem: the upper bound $\bar{\sigma}_{wc}$ of the worst-case standard deviation  $\sigma_{wc}$ is considered as a decision variable and the objective is to minimize $\bar{\sigma}_{wc}$ while meeting the constraint:
\begin{equation}\label{eq:const}
\max_{\boldsymbol\Theta \in \mathcal D_{\boldsymbol \Theta}}\left\|\frac{\mathbf H(\mathrm s,\boldsymbol\Theta)}{\bar{\sigma}_{wc}}\right\|_2\le 1\;.
\end{equation}
Note that on such a robust control design problem, \texttt{systune} provides also a worst-case parametric configuration $\boldsymbol\Theta_{wc}$ saturating the constraint \eqref{eq:const} which can be used to compute a lower bound $\underline\sigma_{wc}$ on $\sigma_{wc}$:
\[
\underline\sigma_{wc}=\|\mathbf H(\mathrm s,\boldsymbol\Theta_{wc})\|_2\:.
\]
This approach is implemented in the function \texttt{wcvariance}. The same approach can be used to compute the worst-case gain $H_\infty$ norm of  $\mathbf H(\mathrm s,\boldsymbol\Theta)$ and implemented in the function \texttt{wcpeak} (the assumption on the strictly properness of $\mathbf H$ can then be relaxed).
Finally the worst-case low frequency DC gain can be computed by computing the worst-case gain at null frequency.  

\section{\textsc{Pelib} overview}
\label{sec:peliboverview}

\begin{wrapfigure}[30]{r}{0.6\columnwidth}
	\begin{center}
		\vspace{-20pt}
		\includegraphics[width=.6\columnwidth]{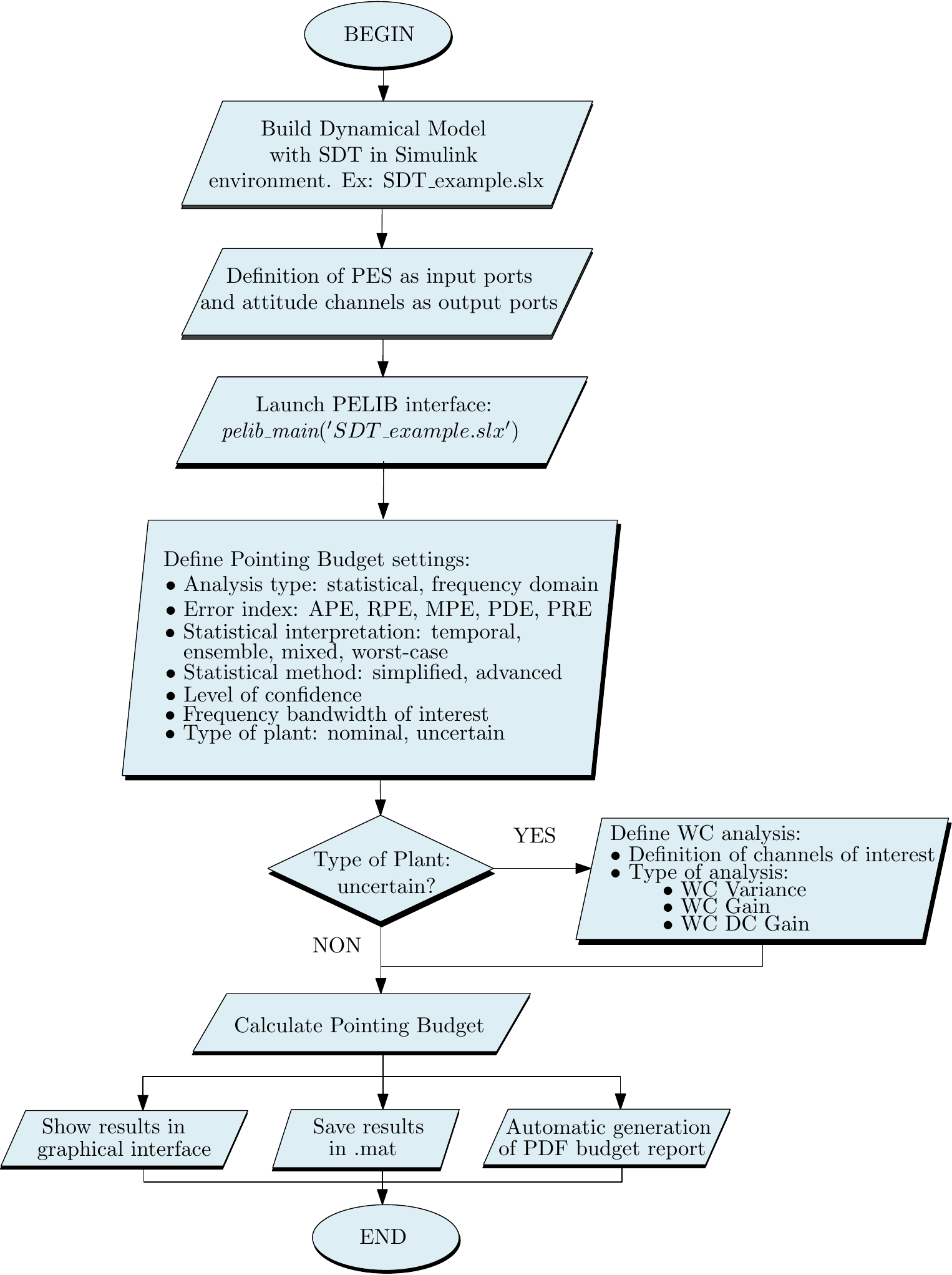}
	\end{center}
	\vspace{-10pt}
	\caption{\textsc{Pelib} analysis flowchart}
	\label{fig:flowchart}
\end{wrapfigure}

\begin{figure}[th!]
\centering
\begin{tabular}{cc}
\subfloat{\includegraphics[width=0.3 \columnwidth]{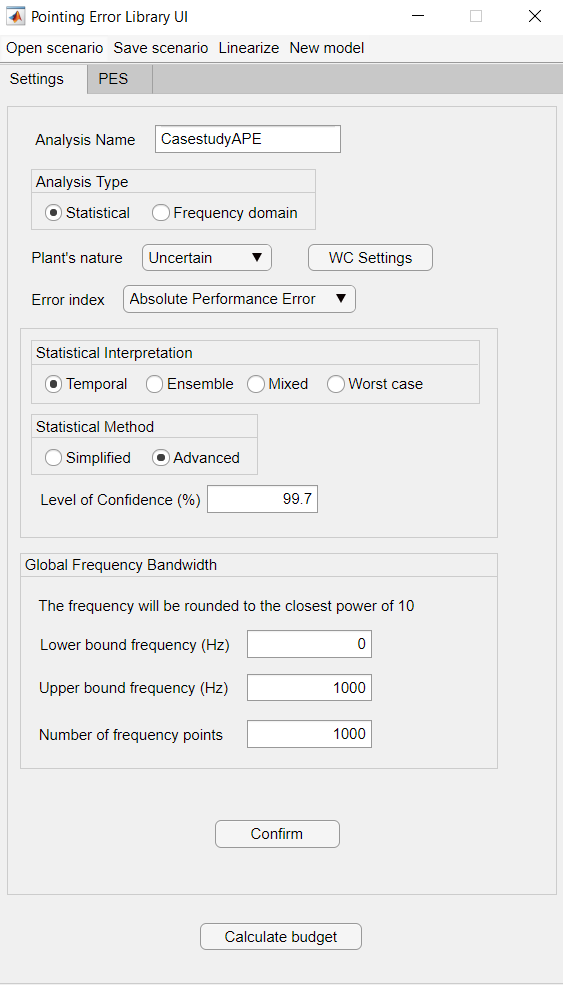}}& 
\subfloat{\includegraphics[width=0.43 \columnwidth]{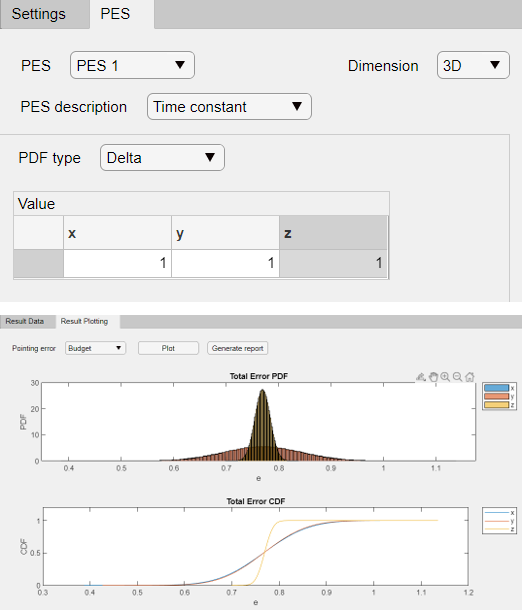}} \\
\end{tabular}
\caption{User interface example: scenario definition and analysis results}
\label{fig:uiexample}
\end{figure}

The flowchart of the main steps to be followed to make a pointing budget with \textsc{PELIB} is depicted in Fig. \ref{fig:flowchart}.
The first step is to build the spacecraft dynamical systems with SDT in Simulink environment and define all PES as input ports and the pointing indexes as output ports. Then a graphical user interface (see Fig. \ref{fig:uiexample}), launched with the command \texttt{pelib\_main(filename)}, allows the user to define the analysis parameters. 

\begin{figure}[h!]
\centering
\includegraphics[width=.3 \columnwidth]{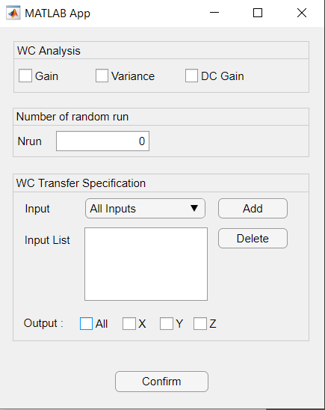}
\caption{Worst-case settings interface}
\label{fig:wcsettingpelib}
\end{figure}

Requirements can be specified in analysis settings tab which will be the first interface the user will see after the tool is launched. In this tab, the specification for the requirements as discussed in section \ref{subsec:requirements} can be defined for the analysis. For accessing the worst-case analysis settings, the user can specify that the system is uncertain and another interface will show up as illustrated by Fig. \ref{fig:wcsettingpelib}. Each PES can be specified on the second tab where the user can choose from the PES database dropdown which consists of all PES listed in Section \ref{subsec:pesdef}.

To display the results, a separate window will be launched after the calculation is finished. There are two tabs to show the results, the first tab lists the results data in \matlab table format. While the second tab is used for visualization of the results depending on their type: the plot of PDF for random variable, PSD for random process or cumulative distribution function (CDF) for the total error budget. In addition to the results, the settings and the input parameters specified previously will also be written as a struct variable on the workspace.

One of the main advantages proposed by this tool is to directly build the spacecraft model and analyze it in the same environment. This is very important for control engineers in order to easily tune and modify the control architecture while meeting all the mission requirements. Another important novelty compared to existing tools is the possibility to take into account parametric and non-parametric uncertainties in the model thanks to the LFT formalism adopted by the SDT \cite{sanfe2019,d_alazard_f_sanfedino_satellite_2021} and then analyze their impact to the pointing performance of the spacecraft with \textsc{Pelib}.

\section{Application}
\label{sec:application}
\begin{figure}[h!]
\centering
\includegraphics[width=.7 \columnwidth]{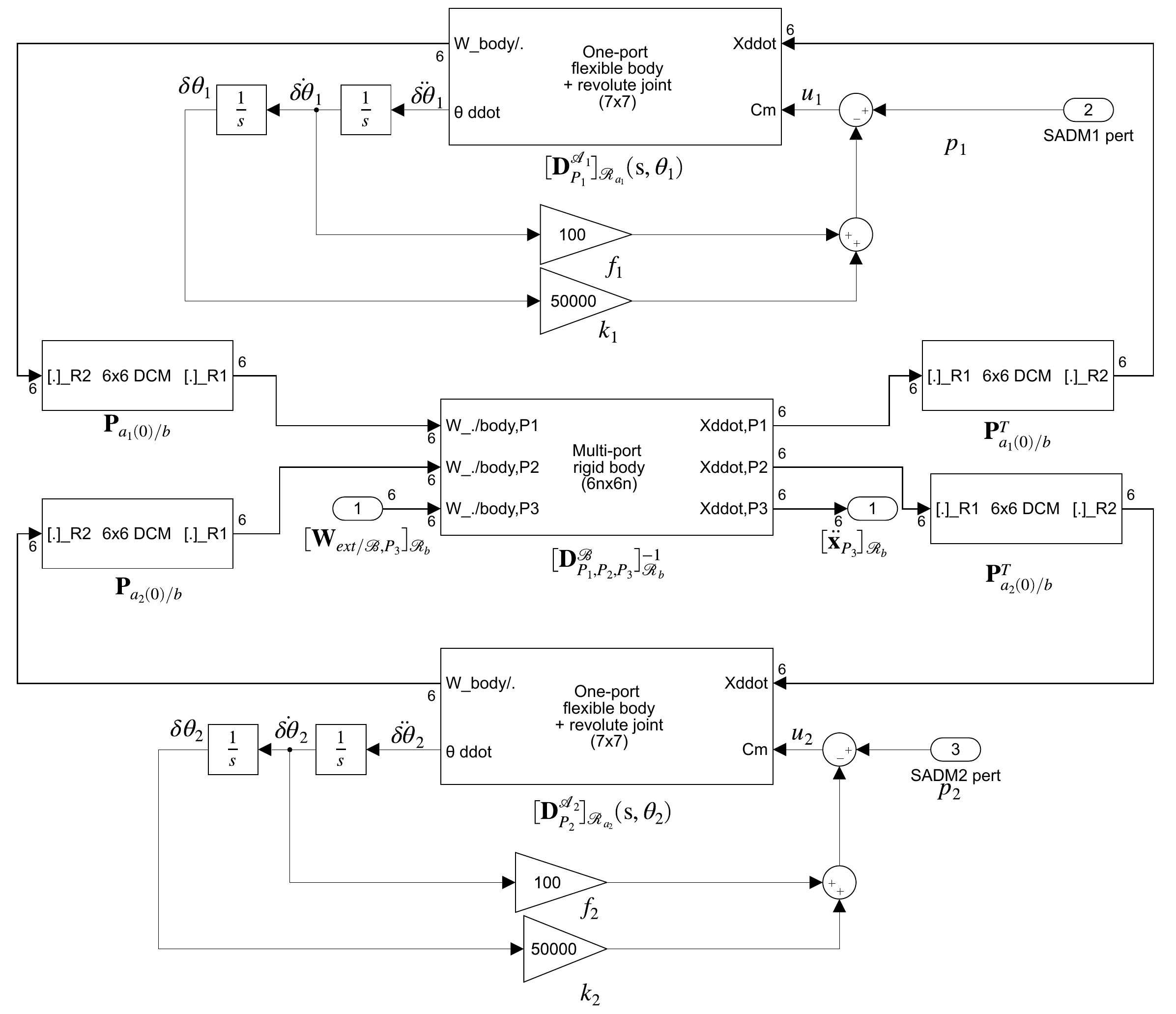}
\caption{Case study: Spacecraft model subsystem using SDT}
\label{fig:simmodel}
\end{figure}

A case study is proposed to further illustrate the capabilities of \textsc{Pelib}. The pointing error budget of a spacecraft with two flexible solar panels driven by SADM is evaluated. The spacecraft dynamical model is built using SDT library in \simulink as shown in Fig. \ref{fig:simmodel}. The parameters used for the spacecraft model are shown in Table \ref{tab:spacecraft_data_sadm}

\begin{table}[h!]
	\renewcommand{\arraystretch}{1.3}
	\caption{Spacecraft data}
	\label{tab:spacecraft_data_sadm}
	\centering
	\begin{footnotesize}
		\resizebox{0.9\textwidth}{!}{\begin{tabular}{ccccc}
				\toprule
				\textbf{System} & \textbf{Parameter} &
				\textbf{Description} & \textbf{Nominal Value } & \textbf{Uncertainty} \Tstrut\Bstrut \\ \toprule
				& $m^{\mathcal{S}}$ & Mass & $1000\,\mathrm{kg}$ & $\pm 20\%$\\
				& $\left[\begin{array}{ccc}
					I_{xx}^{\mathcal{S}} & I_{xy}^{\mathcal{S}} &
					I_{xz}^{\mathcal{S}} \\
					& I_{yy}^{\mathcal{S}} & I_{yz}^{\mathcal{S}} \\
					sym & & I_{zz}^{\mathcal{S}}
				\end{array}\right]$ & Inertia in $\mathcal{R}_s$ frame
				& $\left[\begin{array}{ccc}
					75 & 1 & 2 \\
					& 40 & -1 \\
					sym & & 80
				\end{array}\right]\mathrm{kg\, m^2}$ & $\pm  20\%$\\
				& $\mathbf{r}_G$ & Spacecraft CG &
				$\left[0.35\,\,1.5\,\,0.5\right]\,\mathrm{m}$ & -\\
				\multirow{-6}{*}{\shortstack{Central Body
						\\$\mathcal{S}$}}\\ \hline
				& $m^{\mathcal{A}}$ & Mass & $43\,\mathrm{kg}$ & -\\
				&  $\left[\begin{array}{ccc}
					I_{xx}^{\mathcal{A}} & I_{xy}^{\mathcal{A}} &
					I_{xz}^{\mathcal{A}} \\
					& I_{yy}^{\mathcal{A}} & I_{yz}^{\mathcal{A}} \\
					sym & & I_{zz}^{\mathcal{A}}
				\end{array}\right]$ & Inertia in $\mathcal{R}_a$ frame
				&  $\left[\begin{array}{ccc}
					17& 0 & 0 \\
					& 62 & 0 \\
					sym & & 79
				\end{array}\right]\mathrm{kg\, m^2}$ & -\\
				& $\mathbf{r}_{o}^{\mathcal{A}}$ & $\mathcal{A}$ CG in
				$\mathcal{R}_a$ frame &
				$\left[2.07\,\,0\,\,0\right]\,\mathrm{m}$\\
				& $\tan (\frac{\theta_{r}}{4})$ & Angular configuration & 0 & $\pm 1$\\
				& $\left[\omega_1^{\mathcal{A}}\,
				\omega_2^{\mathcal{A}}\, \omega_3^{\mathcal{A}}\right]$ & Flexible
				modes' frequencies & $\left[5.6\,\, 19.3\,\,35.4\right]\,\mathrm{rad/s}$ & $\pm 20\%$\\
				& $\zeta_1^{\mathcal{A}}, \zeta_2^{\mathcal{A}},
				\zeta_3^{\mathcal{A}}$ &
				Flexible modes' damping & $0.005$ & -\\
				\multirow{-10}{*}{\shortstack{Solar \\ Array
						$\mathcal{A}$}} & $\mathbf{L}_P^{\mathcal{A}}$ & Modal participation
				factors & $\left[\begin{array}{ccc}
					0 & 0 & 0 \\
					0 & 0 & 0 \\
					-5.12 & 0 & -2.97 \\
					0 & -3.84 & 0 \\
					12.5 & 0 & 2.51 \\
					0 & 0 & 0
				\end{array} \right]^{\mathrm{T}}$ & -\\ \hline
		\end{tabular}}
	\end{footnotesize}
\end{table}

\begin{figure}[h!]
	\centering
	\includegraphics[width=0.9\columnwidth]{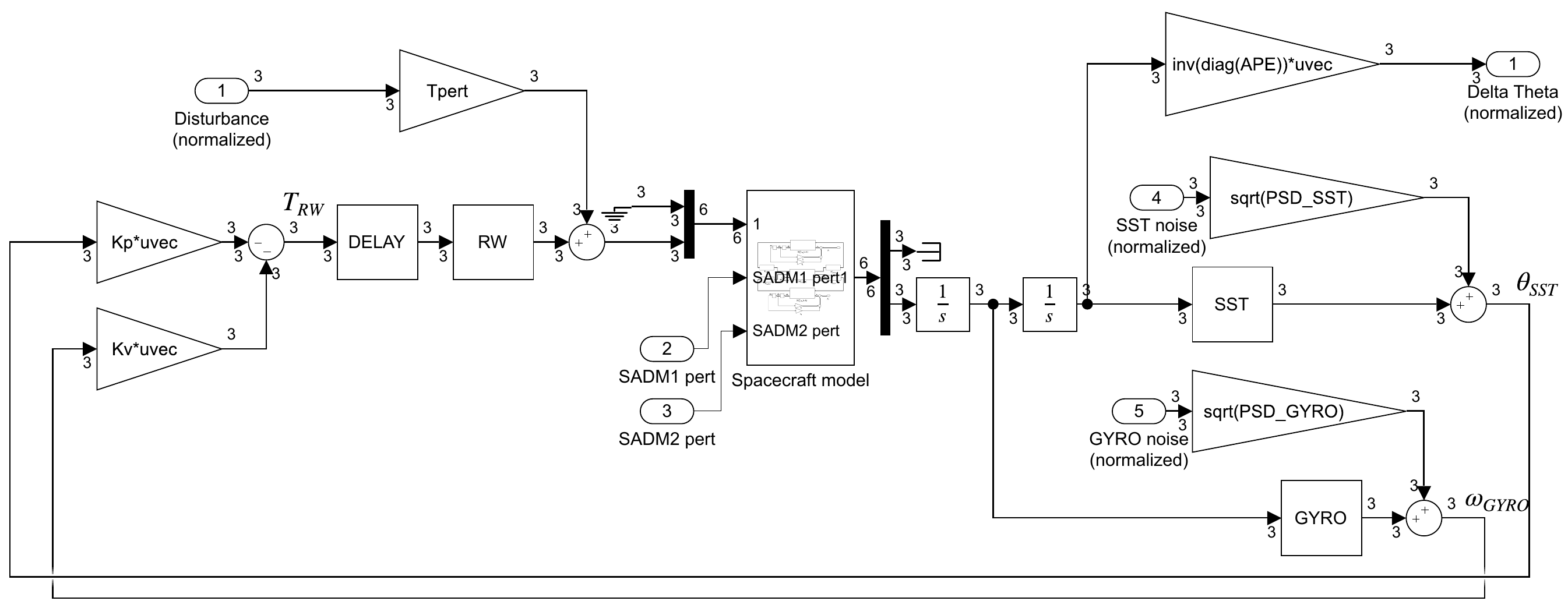}
	\caption{Case study:  \simulink model of a spacecraft with two flexible appendages and five pointing error sources}
	\label{fig:simmodel2}
\end{figure}

The AOCS consists in: a Reaction Wheel Assembly (RWA) modeled as a second order pass filter with a bandwidth of $100\,\mathrm{Hz}$; a star tracker and a three-axis gyro both modeled as a first order low-pass filter with respectively $8\,\mathrm{Hz}$ and $200\,\mathrm{Hz}$ cut-off frequency.
The spacecraft is stabilized using a three-axes decoupled proportional-derivative (PD) control law as illustrated in the closed-loop diagram in Fig. \ref{fig:simmodel2}. The commanded torques for the reaction wheel $T_{RW}$ are given by the equation:
\begin{equation}
	\mathbf{T}_{RW} =  - \mathbf{K}_{p} \bm{\theta}_{\mathrm{SST}} - \mathbf{K}_{v} \bm{\omega}_{\mathrm{GYRO}}
	\label{eq:Controller}
\end{equation} 

The control gains $\mathbf{K}_{p} = \mathrm{diag}({K}_{p_1},{K}_{p_2},{K}_{p_3})$ and $\mathbf{K}_{v}=\mathrm{diag}({K}_{v_1},{K}_{v_2},{K}_{v_3})$ are tuned based on the spacecraft inertia with the relation: ${K}_{p_i} = {I}_{sc_{i}}\cdot{\omega}_{des_i}^2$ and ${K}_{v_i} = {I}_{sc_i}\cdot2\zeta_{des_i}\omega_{des_i}$, with $i=1,2,3$. Here $\omega_{des_i}$, $\zeta_{des_i}$, and $I_{sc_i}$ are the desired closed-loop bandwidth, damping ratio and the spacecraft inertia of the $i-$ axis respectively. For this case study, the desired damping ratio is chosen as 0.7 for all three axes, while the desired bandwidth is calculated based on the spacecraft inertia, APE requirements, and the magnitude of the orbital disturbance given by the equation:
\begin{equation}
	\omega_{des_i} =  \gamma T_{pert}/(I_{sc_i}\theta_{APE_i})
	\label{eq:Controller2}
\end{equation}
where $\gamma$ is the performance margin set to 1.3. 

As shown in Fig. \ref{fig:simmodel2} there are five inputs PES defined in this analysis that correspond respectively to orbital disturbance, star sensor noise, gyro noise and the harmonic disturbances of the two SADM. The output of this model is the angular deviation  $\delta \bm{\theta}$ normalized with the pointing error requirements.  The statistical parameters for each PES are shown in Table \ref{tab:pesstat}.

\begin{table}[h!]
	\caption{Properties of the five pointing error sources considered in the case study}
	\label{tab:pesstat}
	\centering
	\resizebox{1\columnwidth}{!}{
		\begin{tabular}{ccccccccc}
			\toprule
			\textbf{Error Source} & \textbf{Type} & \textbf{PDF} & \textbf{Parameters} & \textbf{X} & \textbf{Y} & \textbf{Z} & \textbf{Unit} \Bstrut \\ \toprule
			Orbital disturbances & Time constant & Delta & Value & 0.03 & 0.01 & 0.02 & Nm \Tstrut\Bstrut \\
			\hline
			\multirow{2}{*}{SADM1 Disturbance} & \multirow{2}{*}{Periodic} & \multirow{2}{*}{Bimodal} & Amplitude & - & - & 0.1 & Nm \\
			&  & & Frequency & - & - & 3.8 & Hz \\
			\hline
			\multirow{2}{*}{SADM2 Disturbance} & \multirow{2}{*}{Periodic} & \multirow{2}{*}{Bimodal} & Amplitude & - & - & 0.1 & Nm\\
			&  & & Frequency & - & - & 3.8 & Hz\\
			\hline
			Star sensor noise & Random process & Gaussian & PSD & 1.$10^{-8}$ & 1.$10^{-8}$ & 1.$10^{-8}$ & $\mathrm{rad}^2/\mathrm{Hz}$ \Tstrut \Bstrut\\
			\hline
			Gyro noise & Random process & Gaussian & PSD & 1.$10^{-10}$ & 1.$10^{-10}$ & 1.$10^{-10}$ & $(\mathrm{rad/s})^2/\mathrm{Hz}$ \Tstrut \\ 
			\bottomrule
		\end{tabular}
	}
\end{table}

\begin{figure}[h!]
\centering
\includegraphics[width=0.7\columnwidth]{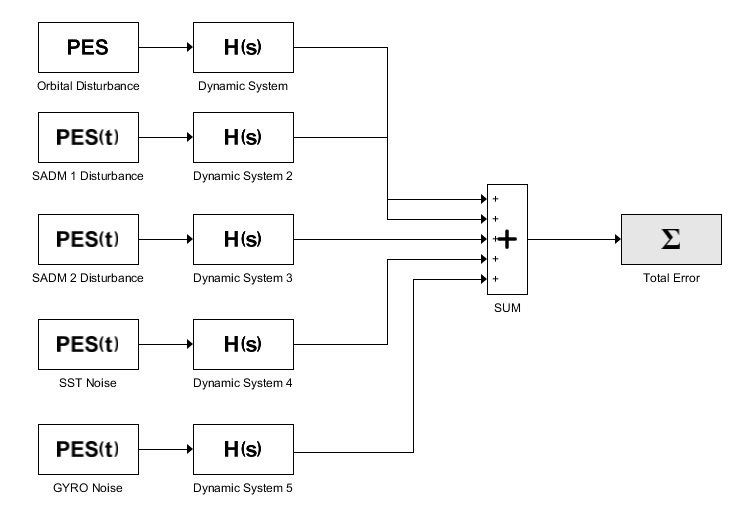}
\caption{Case study:  Pointing system using PEET}
\label{fig:peetmodel}
\end{figure}

In order to validate the pointing budget provided by \textsc{PELIB}, the transfers functions from each PES to the pointing index $\delta\bm{\theta}$ in Fig. \ref{fig:simmodel2} are used to build an equivalent pointing scenario in PEET as shown in Fig. \ref{fig:peetmodel}. 

A second analysis is run as well by considering this time some parametric uncertainties in the spacecraft model (the frequencies of the solar array flexible modes, the mass and the inertia of the central body) according to Table \ref{tab:spacecraft_data_sadm}. The angular configuration $\delta\theta_r$ of the two solar panels (driven by the two SADMs) is also considered as a real uncertainty as shown in \cite{sanfe2019}. Worst-case pointing budget is thus provided together with the corresponding critical values of the uncertain parameters for three case: worst-case variance, worst-case gain and worst-case DC gain.

The detailed settings of the two proposed pointing budgets is outlined in Table \ref{tab:casestudysettings}. 

\begin{table}[h!]
    \caption{Analysis settings}
    \label{tab:casestudysettings}
    \centering
    \resizebox{0.7\columnwidth}{!}{
    \begin{tabular}{ccc}
    \toprule
    \textbf{Parameter} & \textbf{Analysis I} & \textbf{Analysis II} \\
    \toprule
    Error index & APE & RPE \\
    Time window & - & 3 ms \\
    Pointing error requirement & [0.1745 0.1745 0.873] mrad & [0.06 0.06 0.06] $\mu$rad \\
    Statistical interpretation & Temporal & Temporal \\
    Statistical Method & Advanced & Advanced \\
    Level of confidence & 99.7\% & 99.7\%\\
    Type of plant & Nominal & Uncertain \\
    \hline
    \end{tabular}}
\end{table}
\section{Results and discussion}
\label{sec:result}

\subsection{Nominal system analysis}

The results for the nominal error analysis (Analysis I in Table \ref{tab:casestudysettings}) are presented on Table \ref{tab:statresnom} and Table \ref{tab:budgetresnom}. Table \ref{tab:statresnom} shows the comparison of the statistics of each error source after AST-2 and AST-3 between PEET and \textsc{Pelib} and their relative error.

\begin{table}[h!]
	\caption{Error contributions comparison after transfer analysis}
	\label{tab:statresnom}
	\centering
	\resizebox{1\textwidth}{!}{\begin{tabular}{ccccccc}
			\toprule
			\multicolumn{1}{c}{\multirow{2}{*}{\textbf{Error source}}} &
			\multicolumn{2}{c}{\textbf{PEET}} &
			\multicolumn{2}{c}{\textbf{PELIB}} & 
			\multicolumn{2}{c}{\textbf{Deviation(\%)}} \\
			\cmidrule(lr){2-3}
			\cmidrule(lr){4-5}
			\cmidrule(lr){6-7}
			&
			\multicolumn{1}{c}{Mean} &
			\multicolumn{1}{c}{Standard deviation} &
			\multicolumn{1}{c}{Mean} &
			\multicolumn{1}{c}{Standard deviation} &
			\multicolumn{1}{c}{Mean} &
			\multicolumn{1}{c}{Standard deviation}\\
			\midrule[\heavyrulewidth]
			Orbital disturbance & [0.7692 0.7692 0.7692] & [0 0 0] & [0.7692 0.7692 0.7692] & [0 0 0] & [0 0 0] & [0 0 0] \\
			SADM1 disturbance & [0 0 0] & [2.74 0.546 81.3].$10^{-5}$ & [0 0 0] & [2.74 0.546 81.3].$10^{-5}$ & [0 0 0] & [0 0 0] \\
			SADM2 disturbance & [0 0 0] & [2.74 0.546 81.3].$10^{-5}$ & [0 0 0] & [2.74 0.546 81.3].$10^{-5}$ & [0 0 0] & [0 0 0]\\
			Star sensor noise & [0 0 0] & [9.95 7.55 1.95].$10^{-2}$ & [0 0 0] & [9.95 7.55 1.95].$10^{-2}$  & [0 0 0] & [0 0 0]\\
			Gyro noise & [0 0 0] & [2.63 3.46 0.537].$10^{-2}$ & [0 0 0] & [2.64 3.47 0.537].$10^{-2}$ & [0 0 0] & [0.38 0.29 0]\\
			\bottomrule
	\end{tabular}}
\end{table}

Table \ref{tab:budgetresnom} summarizes the budget for each error type as well as the final error budget. The error budgets obtained using PEET and \textsc{Pelib} are shown to be almost identical with a maximum deviation of 0.2435\%. The largest contribution to the APE error budget comes from the time constant PES which corresponds the orbital disturbances. On the other hand the contribution of periodic PES is negligible for the APE requirement. The total error budget for the nominal case for both PEET and \textsc{Pelib} are shown to be below the requirement for $y$ and $z$-axes. However the APE error slightly exceeds the requirement on $x$-axis by $\approx 5\%$. Moreover it has to be noticed that for $y$-axis, there is a very small margin ($\approx 2\%$) which can easily exceeded when uncertainties are included.

\begin{table}[h!]
    \caption{Error budget comparison for nominal system}
    \label{tab:budgetresnom}
    \centering
    \resizebox{0.9\textwidth}{!}{
    \begin{tabular}{cccccccccc}
    \toprule
    \multicolumn{1}{c}{\multirow{2}{*}{\textbf{Error budget}}} &
    \multicolumn{3}{c}{\textbf{PEET}} &
    \multicolumn{3}{c}{\textbf{PELIB}} &
    \multicolumn{3}{c}{\textbf{Deviation (\%)}}\\ 
    \cmidrule(lr){2-4}
    \cmidrule(lr){5-7}
    \cmidrule(lr){8-10}
         &
    \multicolumn{1}{c}{X} &
    \multicolumn{1}{c}{Y} & 
    \multicolumn{1}{c}{Z} &
    \multicolumn{1}{c}{X} &
    \multicolumn{1}{c}{Y} & 
    \multicolumn{1}{c}{Z} &
    \multicolumn{1}{c}{X} &
    \multicolumn{1}{c}{Y} &
    \multicolumn{1}{c}{Z} \\
    \midrule[\heavyrulewidth]
    Time constant & 0.7692 & 0.7692 & 0.7692 & 0.7692 & 0.7692 & 0.7692 & 0 & 0 & 0\\
    Random process & 0.3055 & 0.2464 & 0.06 & 0.3053 & 0.2458 &	0.06 & 0.0655 & 0.2435 & 0\\
    Periodic & 7.75E-05 & 1.54E-05 & 0.0023 & 7.75E-05 & 1.54E-05 & 0.0023 & 0 & 0 & 0\\
    Total & 1.052 &	0.9977 & 0.8248 & 1.0535 & 0.9976 &	0.825 & 0.143 & 0.01 & 0.0242\\
    \bottomrule
    \end{tabular}}
\end{table}

\subsection{Worst-case analysis with uncertain plant}

In this section the worst-case analysis (Analysis II in Table \ref{tab:casestudysettings}) carried out using the RPE metric on the case study will be discussed. 

\begin{figure}[h!]
\centering
\includegraphics[width=1\columnwidth]{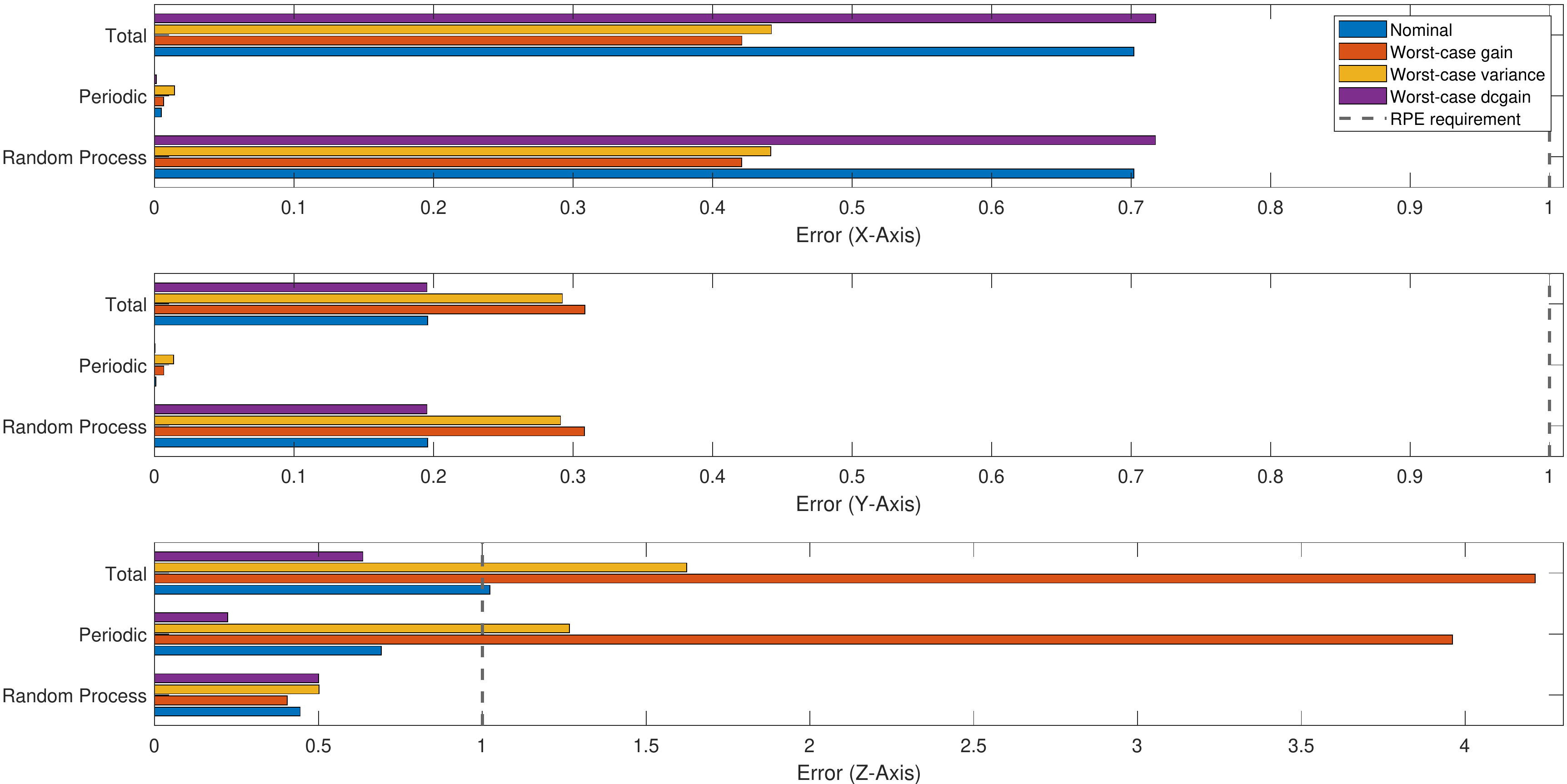}
\caption{Error budget comparison for worst-case analysis}
\label{fig:budgetwc}
\end{figure}

The error budget comparison can be seen in Fig. \ref{fig:budgetwc}. For RPE metric, compared to the APE requirement, periodic error sources have quite significant contribution to the nominal system error budget. This can be expected since the RPE requirement is significantly more stringent compared to the APE around the frequency of the SADM disturbance. It can be observed as well that, for the nominal system, the error budgets are far below the requirement for $x$ and $y$-axis, while on $z$-axis, the RPE value already slightly exceeds the limit of 0.06 $\mu$rad. Notice that $z$-axis corresponds to the SADM rotor axis. 

The introduction of uncertainties in the system makes the RPE requirement on $z$-axis to be largely broken by exceeding of more than four times the design limit in the total budget, while the worst-case for $x$ and $y$-axes meet always the pointing design in all scenarios.
 
As seen for the nominal case, also with an uncertain plant, the major contribution for the worst-case RPE budget on the $z$-axis comes from the periodic disturbances of two SADMs.

\begin{figure}[h!]
	\centering
	\includegraphics[width=\columnwidth]{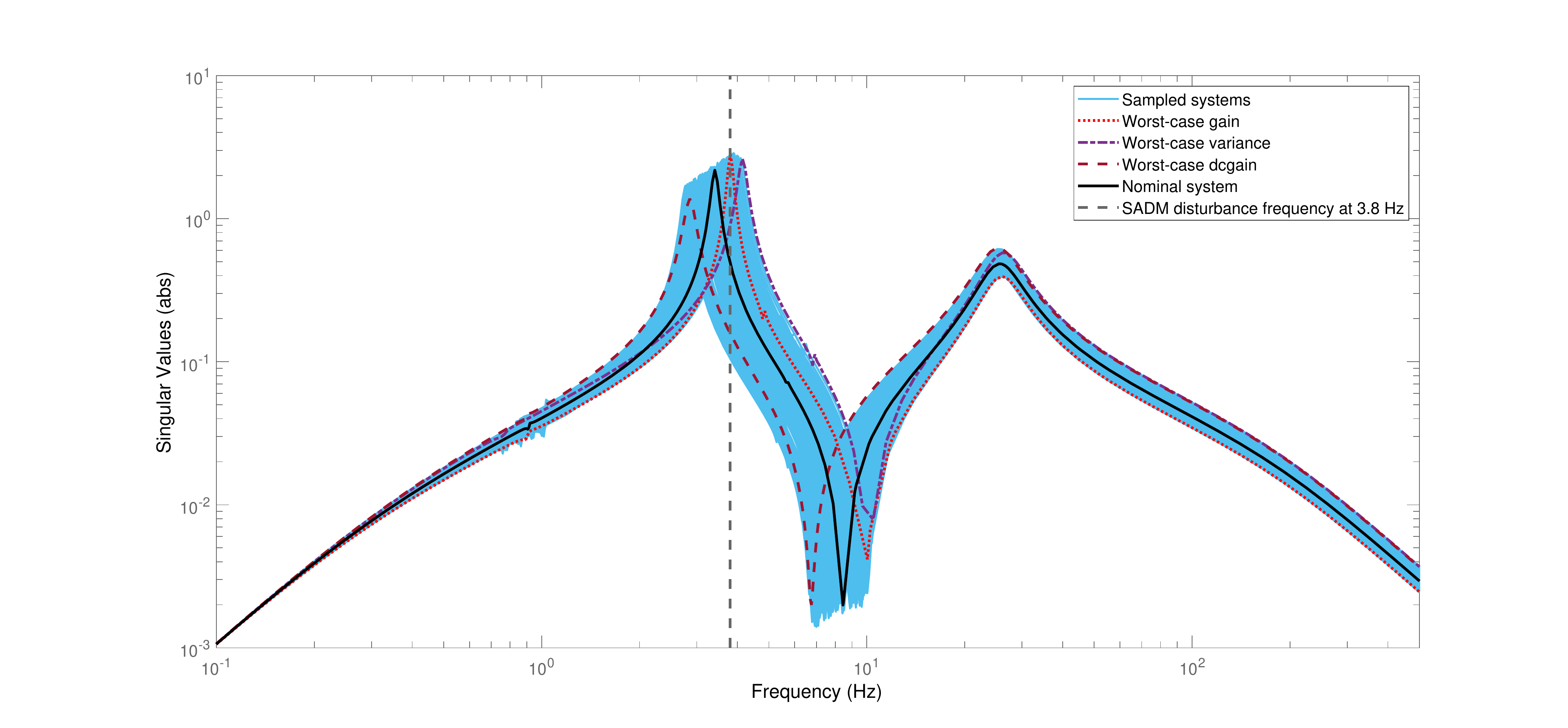}
	\caption{Singular values of the transfer function from the two SADM disturbances to the $z$-axis RPE performance channel, normalized with the maximum SADM disturbance amplitude}
	\label{fig:svsadm}
\end{figure}

The impact of the SADM disturbance on the RPE is highlighted in Fig. \ref{fig:svsadm}. This figure shows the singular values of the transfer between the two SADM disturbances to the $z$-axis RPE performance channel normalized with the maximum SADM disturbance amplitude. It can be noticed how for the worst-case gain scenario one of the spacecraft flexible mode corresponds exactly to the SADM disturbance frequency (3.8 Hz), that explains the results previously discussed in Fig. \ref{fig:budgetwc}.

Table \ref{tab:wcconfig} finally shows the worst-case parameter configurations for each of the three worst-case analyses.

\begin{table}[h!]
	\caption{Worst-case configurations for each criterion}
	\label{tab:wcconfig}
	\centering
	\resizebox{0.9\columnwidth}{!}{
		\begin{tabular}{ccccccc}
			\hline
			\textbf{Parameter} & \textbf{Symbol} & \textbf{Unit} & \textbf{Nominal} & \textbf{WC Gain} & \textbf{WC Variance} & \textbf{WC DC Gain} \Tstrut\Bstrut \\
			\hline
			Central body mass & $m^{\mathcal{S}}$ & kg & 1000 & 1165 & 800 & 800  \Tstrut\Bstrut \\
			& $I_{xx}$ &  & 75 & 76.7261 & 60 & 61.82 \\
			Central body inertia & $I_{yy}$ & $\mathrm{kg.m^{2}}$ & 40 & 48 & 32 & 46.78 \\
			& $I_{zz}$ &  & 80 & 96 & 64 & 64 \\
			SA angular configuration & $\mathrm{tan}(\theta_r /4)$ & - & 0 & 0.33 & 0.55 & 0.007 \\
			& $\omega_{1}$ &  & 5.6 & 5.42 & 4.48 & 5.53 \\
			Flexible mode's frequencies & $\omega_{2}$ & $\mathrm{rad/s}$ & 19.3 & 22.37 & 23.16 & 15.44 \\
			& $\omega_{3}$ &  & 35.4 & 29.96 & 42.48 & 41.96 \\
			\hline
	\end{tabular}}
\end{table}

\section{Conclusion}
This paper proposed a novel tool to perform pointing performance analysis for space missions following the European standards. \textsc{Pelib} is designed as an extension of the SDT already developed in the past for deriving minimal LFT models of flexible space structures. \textsc{Pelib} allows the user of SDT to perform pointing performance analyses of real mission scenarios within the same environment used for control synthesis and spacecraft dynamics simulation. Two methods have been proposed to compute worst-case variance and worst-case gain in order to make possible pointing analyses in \textsc{Pelib} by including both complex and parametric uncertainties in the system. This feature represents an enhancement in the pointing V\&V process with respect to the current reference tools. Only one analysis is in fact needed to cover the whole set of system uncertainties and provide the worst-case pointing performance without relying on time-consuming and no-global approaches like Monte Carlo campaigns as classically proposed in V\&V activities of aerospace industries.

The capabilities of the proposed tool were demonstrated in a case study involving a spacecraft model with two flexible solar arrays and SADMs. The analysis for nominal system was compared with the results obtained with PEET, the European reference tool for error budgeting used in several ESA missions. It was shown that \textsc{Pelib} managed to obtain the same results as PEET in the nominal case. The case study further showed \textsc{Pelib} capabilities in providing pointing budget for three worst-case scenarios: worst-case variance, worst-case gain and worst-case DC gain. The combination of critical values of mechanical parameter are provided as well for the three cases.

\bibliographystyle{IEEEtran}
\bibliography{biblist}

\end{document}